\documentclass[aps,prl,twocolumn,showpacs,superscriptaddress]{revtex4}
\usepackage{amsmath}
\usepackage{amsfonts}
\usepackage{amssymb}
\usepackage{graphicx}
\usepackage{epsfig}

\begin{document}

\title{Metastability, Nucleation, and Noise-Enhanced Stabilization Out of Equilibrium}
\author{Pablo I. Hurtado}
\affiliation{Laboratoire des Collo\"ides, Verres et Nanomat\'eriaux, Universit\'e
Montpellier II, Montpellier 34095, France}
\author{J. Marro}
\author{P.L. Garrido}
\affiliation{Institute \emph{Carlos I} for Theoretical and Computational Physics, and
Departamento de Electromagnetismo y F\'{\i}sica de la Materia, Universidad
de Granada, E-18071-Granada, Espa\~{n}a}

\begin{abstract}
We study metastability and nucleation in a kinetic 2D--Ising model which is driven out of
equilibrium by a small random perturbation of the usual dynamics at 
temperature $T$. We show that, at a mesoscopic/cluster level, a
nonequilibrium potential describes in a simple way
metastable states and their decay. We thus predict noise--enhanced stability 
of the metastable phase and resonant propagation of domain walls at low--$T$. 
This follows from the nonlinear interplay between thermal and nonequilibrium fluctuations, 
which induces reentrant behavior
of the \emph{surface tension} as a function of $T$. Our results, which are confirmed by Monte Carlo
simulations, can be also understood in terms of a Langevin equation with competing additive
and multiplicative noises.
\end{abstract}

\maketitle

Relaxation in many natural systems proceeds through metastable states 
\cite{Penrose,Gunton,Rikvold,Debenedetti,Stillinger,Strumia,protein,resonant,FDT}.
This is often observed in condensed matter physics \cite{Stillinger}, and also in various
other fields, from cosmology \cite{Strumia} to biology \cite{protein} and high-energy physics \cite{resonant}.
In spite of such ubiquity, the microscopic understanding of metastability still
raises fundamental questions. A main difficulty is that this is a dynamic
phenomenon not included in the ensemble formalism. Even so, metastable 
states may be regarded in many cases as equilibrium states for times short 
compared with their relaxation time, and one may derive macroscopic properties 
from \emph{restricted ensembles} \cite{Penrose}, or obtain fluctuation--dissipation 
theorems \cite{FDT}. On the long run, however, metastable states eventually decay 
triggered by rare fluctuations. This relaxation can be described in terms of free-energy 
differences as far as one is dealing with systems evolving towards equilibrium steady
states \cite{Rikvold}. However, as a rule, natural systems are open to the environment, 
which induces currents of macroscopic observables or competitions between opposing 
tendencies which typically break detailed balance \cite{MarroDickman}. Consequently, 
in many cases, stationary states are not equilibrium states, but are strongly influenced 
by dynamics, which adds further challenge to the microscopic understanding of metastability.

In this paper we report on the nature of metastability in a full
nonequilibrium setting. In order to focus on the basic physics, 
we study the simplest nonequilibrium model in which metastable states are
relevant, namely an Ising model with dynamic impurities. 
We show that introducing a nonequilibrium condition has dramatical effects on 
metastable dynamics. In particular, we find noise-enhanced stabilization of metastable 
states, resonant propagation of domain walls and other novel low temperature physics
not observed in equilibrium. Surprisingly, these nonequilibrium phenomena can be understood 
via extended nucleation theory, starting from a nonequilibrium potential or \emph{free--energy} at a 
mesoscopic/cluster level. This is possible because the excitations responsible for the 
metastable-stable transition result from the competition between a surface
and a bulk term, as in equilibrium. We thus anticipate that a similar approach can be used to 
understand metastability in many other nonequilibrium systems in which a surface/bulk competition
is a dominant mechanism. 
Our results are also relevant for metastable nanodevices where 
nonequilibrium impurities play a fundamental role \cite{MarroDickman}.

Consider a two-dimensional square lattice of side $L$ with periodic boundary
conditions. We define a spin variable $s_{i}=\pm 1$ at each node, $i\in
\lbrack 1,N\equiv L^{2}].$ Spins interact among them and with an external
magnetic field $h$ via the Ising Hamiltonian function, $\mathcal{H}%
=-\sum_{\langle ij \rangle}s_{i}s_{j}-h\sum_{i}s_{i},$ where
the first sum runs over all nearest--neighbors pairs. We also define a
stochastic single spin--flip dynamics with transition rate 
\begin{equation}
\omega (\mathbf{s}\rightarrow \mathbf{s}^{i})=p+(1-p)\Psi (\beta \Delta 
\mathcal{H}_{i}),  \label{rate}
\end{equation}%
where $\mathbf{s}=\left\{ s_{k}\right\} $ and $\mathbf{s}^{i}$ stand for the
configurations before and after flipping the spin at node $i,$ respectively, 
$\Delta \mathcal{H}_{i}$ is the \textit{energy} increment in such flip, and $%
\beta =1/T.$ The function in (\ref{rate}) is chosen here as 
$\Psi (x)=(1+\text{e}^x)^{-1}$, which corresponds to the \textit{Glauber rate}.
However, similar results hold also for the \emph{Metropolis rate}, and possibly for many other 
local, detailed-balanced dynamical rules $\Psi$.

For any $0<p<1$ two different heat baths compete in (\ref{rate}): 
One is at temperature $T,$ which operates with probability $(1-p)$, 
and the other induces completely random spin--flips (as a bath at 
\emph{infinite} temperature would do) with probability $p.$ As a result of
this competition, a\textit{\ nonequilibrium} steady state sets in
asymptotically \cite{MarroDickman,Vacas,Pablo1,Pablo2}, which cannot be
characterized by any Gibbsian measure. For $h=0,$ the model exhibits an
order--disorder continuous phase transition at temperature 
$T_{c}(p)<T_{c}(p=0)\equiv T_{\text{ons}}$, where $T_{\text{ons}}=2/\ln (1+%
\sqrt{2})$, and the order washes out for any $p>p_{c}=(\sqrt{2}-1)^{2}\approx 0.17,$ 
even at $T=0$. This phase transition belongs to the Ising universality class, a result that 
has lead to the belief that two-temperature nonequilibrium Ising models behave in general as
their equilibrium ($p=0$) counterpart. We show below that, in what concerns metastability, essential 
differences indeed exist.

We remark that our motivation here for the random perturbation $p$ is that it is
the simplest microscopic mechanism that induces nonequilibrium behavior. 
However, similar dynamic impurities play a fundamental role in some natural systems. 
As a matter of fact, an equivalent mechanism has been used to model the
macroscopic consequences of rapidly--diffusing local defects \cite{MarroDickman} 
and quantum tunneling \cite{Vacas} in magnetic samples and,
more generally, the origin of scale-invariant avalanches \cite{aval}.
\begin{figure}[b]
\centerline{
\psfig{file=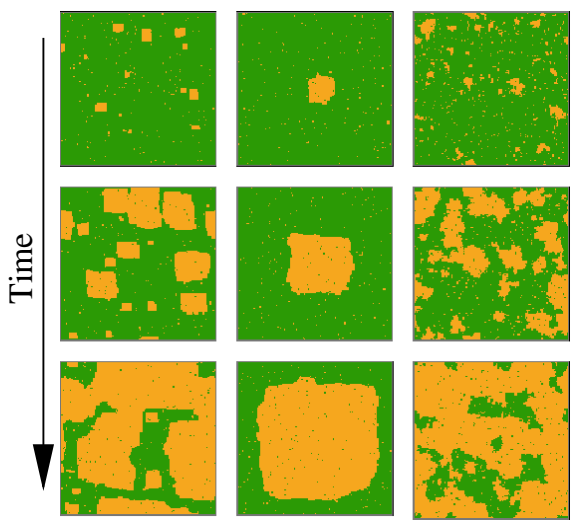,width=7.5cm}}
\caption{{\small (Color online) Escape configurations
for $L=128$, $p=0.01$, $h=-0.25$ and $T/T_{\text{ons}}=$ 0.1, 0.3 and 0.7,
respectively from left to right.}}
\label{DSP0}%
\end{figure}

For small $h<0$ and $T<T_{c}(p),$ an initial homogeneous state with all spins up is
metastable. It will eventually decay$\phantom{s}$ toward the stable state of
magnetization $m=N^{-1}\sum_{i}s_{i}<0$. Inspection of escape configurations
(Fig. \ref{DSP0}) shows that this is a highly inhomogeneous process
triggered by (large) compact clusters of the stable phase. 
These excitations 
then grow or shrink in the metastable sea depending on the competition 
between their surface, which hampers cluster growth, and their bulk, which favors it. 
In equilibrium ($p=0$), this competition is controlled by the cluster
interfacial free--energy, or surface tension. Far from equilibrium ($0<p<1$), despite
the lack of a proper bulk free--energy function, one may define \cite{Pablo2} an
effective \emph{surface tension} $\sigma _{0}(T,p),$ that captures the
properties of the nonequilibrium interface. This is based on the assumption
that the normalization of a probability measure for interface configurations
can be interpreted as a sort of nonequilibrium \textit{partition function }%
\cite{Pablo2}. Similar assumptions have been shown to yield excellent
results for nonequilibrium phase transitions \cite{LY}. Interesting enough, 
$\sigma _{0}(T,p)$ exhibits non--monotonous temperature dependence for any 
$0<p<1$ with a maximum at a non-trivial value of $T$ \cite{Pablo2,sigma}.
\begin{figure}[t]
\centerline{
\psfig{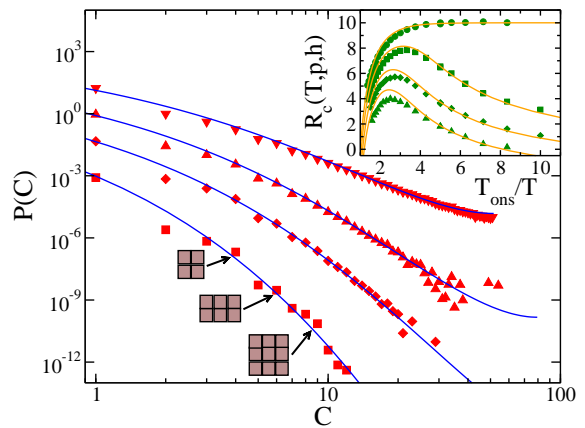}}
\caption{{\small (Color online) Cluster distribution $P(C)$ for $T=0.2T_{\text{ons}}$, 
$L=53$, $h=-0.1$ and, from bottom to top, $p=0.001$, $0.005$, $0.01$, 
$0.02$, with $\protect\gamma =0.815$, $0.82$, $0.83$, $0.85$, 
respectively. Lines are theoretical predictions and points are MC results. 
\protect{Inset}: $\mathcal{R}_{c}$ vs. $T_{\text{ons}}/T$ for $L=53$, 
$h=-0.1$ and, from top to bottom, $p=0$, $0.001$, $0.005$ and $0.01$. The 
$n^{th}$ curve (from bottom to top) has been rescaled by a factor $10^{(n-1)}$ 
(main plot), or shifted $(4-n)$ units (inset) in the $\hat{y}$-axis. }}
\label{Rc}%
\end{figure}

Consequently with the surface/bulk competition driving the relevant excitations,
it is straightforward to write down an ansatz for an
effective \emph{free-energy} cost of a cluster of radius $R$, 
$\mathcal{F}(R)=\gamma \lbrack 2\Omega R\sigma _{0}-\Omega R^{2}2m_{0}|h|]$; 
see \cite{Rikvold,Debenedetti}. 
We may then derive the
zero--field spontaneous magnetization $m_{0}(T,p)$ within a mean--field
approximation \cite{Pablo1} and the cluster form factor $\Omega (T,p)$ from $%
\sigma _{0}$ via the Wulff construction \cite{Pablo2}; $\gamma $ is a
weakly--varying parameter, very close to one, that stands for small
corrections to classical nucleation theory \cite{Debenedetti}. The critical
cluster radius, such that supercritical (subcritical) clusters tend to grow
(shrink), thus follows as $\mathcal{R}_{c}=\sigma _{0}/(2m_{0}|h|).$
Estimates of $\mathcal{R}_{c}$ in MC simulations (see \cite{large} for
details) are shown in the inset to Fig. \ref{Rc} together with the
analytical predictions. The agreement is excellent for temperatures well
below $T_{c}(p)$ and, most important, $\mathcal{R}_{c}(T,p,h)$ exhibits
non-monotonous $T$-dependence for any $p>0.$

Our ansatz above also implies a Boltzmann--like distribution, 
$P(C)=\mathcal{M}^{-1}\exp [ -\beta {\cal F}(\sqrt{C/\Omega})]$ 
for the fraction of stable--phase clusters of volume $C=\Omega
R^{2}$ in the metastable phase. The normalization $\mathcal{M}=2\Theta /(1-m)
$, with $\Theta =\sum_{C=1}^{C_{\ast }}C\exp [-\beta \mathcal{F}(\sqrt{C/\Omega})]$ 
and $C_{\ast }=\Omega \mathcal{R}_{c}^{2}$, is defined so that the metastable
state has the mean--field magnetization $m(T,p,h)$ \cite{Pablo1}. Fig. \ref%
{Rc} depicts our results for $P(C)$. Again, theoretical predictions compare
very well to MC results. For $p=0.001$, MC data reveal a non-trivial
structure in $P(C)$ which is not captured by our continuous theory. This is
due to the lattice structure which, for low-$T$ and small $p,$ gives rise to 
resonances for clusters with complete \emph{shells}, i.e. $2\times 2
$, $3\times 2$, $3\times 3;$ see Fig. \ref{Rc}. For larger $p$ and/or $T$,
fluctuations wash out this effect. Also interesting in Fig. \ref{Rc} is that
the nonequilibrium perturbation $p$ enhances fluctuations and favors larger
clusters.
\begin{figure}[t]
\centerline{
\psfig{file=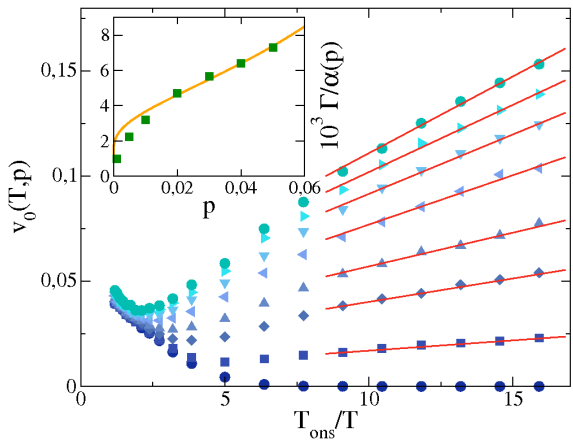,width=7.5cm}}
\caption{{\small (Color online) 
$v_0$ vs. $T_{\textrm{ons}}/T$ for $h=-0.1$ and
$p=0, \ 0.001,  \ 0.005, \ 0.01, \ 0.02, \ 0.03, \ 0.04, \ 0.05$ (from bottom to top). Curves are 
linear fits to data. Inset: Slope of linear fits vs. $p$. The line is the theoretical prediction,
with $\Gamma \approx 0.0077$.}}
\label{vel}%
\end{figure}

The assumed ${\cal F}(R)$ also involves a force per unit area which 
controls the growth of supercritical droplets. In particular, the 
propagation velocity of a large cluster should obey the Allen-Cahn 
expression $v_{0}=2\Gamma ^{\prime }m_{0}|h|/\sigma _{0}$ 
\cite{Gunton}, where $\Gamma ^{\prime }$ is a non-universal constant. Our
estimates for $v_{0}$ in MC simulations of a flat propagating interface (an
infinitely-large cluster) \cite{large} are in Fig. \ref{vel}. The fact
that $v_{0}$ exhibits the expected non-monotonous $T$-dependence means that cooling the
system favors domain wall propagation in a nonequilibrium setting. 
Moreover, $\sigma _{0}\approx \alpha (p)T$ at low--$T$, with $\alpha (p)=\ln [(1-\sqrt{p%
})/(p+\sqrt{p})]$ \cite{Pablo2,sigma,large}, so that we expect $v_{0}\approx
\lbrack \Gamma /\alpha (p)](T_{\text{ons}}/T)$ in this regime, where 
$\Gamma$ is a different non-universal constant and we assumed 
$m_{0}(T\rightarrow 0,p)\approx 1$. This is nicely confirmed in simulations,
see inset to Fig. \ref{vel}. 

The nucleation rate ${\cal I}$ for critical clusters determines the metastable-state lifetime. 
From our hypothesis, 
$\mathcal{I}=A|h|^{\delta }\exp [-\beta \mathcal{F}(\mathcal{R}_{c})]$, where 
$A(p)$ is a non-universal amplitude and $\delta \approx 3$ for random
updatings \cite{Rikvold}. The relaxation pattern depends on the balance between 
two different length scales, namely, $L$ and the mean cluster separation $\mathcal{R%
}_{0}(T,p,h)=(v_{0}/\mathcal{I})^{1/3}$ (typically, $\mathcal{R}_{c}\ll 
\mathcal{R}_{0}$, $L$). For $\mathcal{R}_{0}\gg L$ [Single--Droplet (SD)
regime], nucleation of a \emph{single} critical cluster is the relevant
excitation, and the metastable-state lifetime is $\tau _{\text{SD}}=(L^{2}%
\mathcal{I})^{-1}.$ For $\mathcal{R}_{0}\ll L$ [Multidroplet (MD) regime],
the metastable-stable transition proceeds via the nucleation of many
critical clusters, and $\tau _{\text{MD}}=[\Omega v_{0}^{2}\mathcal{I}/(3\ln
2)]^{-1/3}$ \cite{Rikvold}. The crossover corresponds to the dynamic spinodal point, $|h_{%
\text{DSP}}|(T,p)=\Omega \gamma \sigma _{0}^{2}/(6m_{0}T\ln L),$ such that the
SD (MD) regime holds for $|h|<|h_{\text{DSP}}|$ ($|h|>|h_{\text{DSP}}|$).
\begin{figure}[t]
\centerline{
\psfig{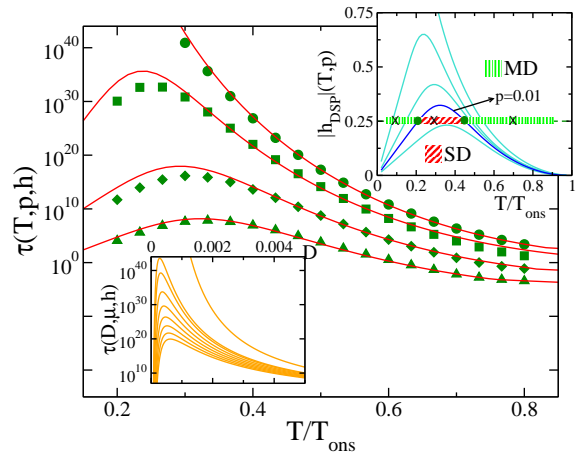}}
\caption{{\small (Color online) Lifetime $\tau $ vs. $T$ for
the same $p$ and conditions than for the inset in Fig. \protect\ref{Rc}. The 
$n^{th}$ curve (from top to bottom) is rescaled by a factor $10^{-2(n-1)}$.
Solid lines are theoretical predictions for (from top to bottom) $\protect%
\gamma =1,\ 0.85,\ 0.77,\ 0.65$. Amplitudes $A(p)\in \lbrack 10^{-3},10^{-2}]$. 
Top inset: $|h_{\text{DSP}}|$ vs. $T$ for (from top to bottom) 
$p=0,\ 0.001,\ 0.005,\ 0.01,\ 0.02$. Bottom inset: $\protect%
\tau $ vs. $D$, as derived from Langevin equation, for $h=-0.1$ and $%
\protect\mu \in \lbrack 0,6\times 10^{-3}]$, increasing from top to bottom.}}
\label{vida}
\end{figure}

We measured $\tau $ in MC simulations using rejection-free methods \cite%
{large,MCAMC}. This is in Fig. \ref{vida} together with our predictions.
Interesting enough, we observe that $\tau $ increases with $T$ for fixed $p>0$ 
at low--$T$. That is, the local stability of \emph{nonequilibrium}
metastable states is enhanced by the addition of thermal noise. This
behavior resembles the noise--enhanced stability (NES) phenomenon reported
in experiments on unstable systems \cite{Mantegna}, which is in
contrast with the simple Arrhenius law observed in equilibrium.
Remarkably, only \emph{thermal} NES is observed: Increasing $p$ for fixed $T$
always results in shorter $\tau $. This complex phenomenology is captured by
our simple ansatz, which traces back the NES phenomenon to the low--$T$
anomaly in $\sigma _{0}$. 
More generally, the stochastic resonance in $\tau$ (and $v_0$) stems from a nonlinear
cooperative interplay between thermal and nonequilibrium noises. Although
both noise sources induce disorder when applied independently, their combined
effect results in a resonant stabilization of the metastable phase at low--$T$. 
This non-linear effect is also reflected in the morphology of the
metastable--stable transition. In particular, $|h_{\text{DSP}}|$ inherits
the non-monotonous $T$-dependence of $\sigma _{0}$, see top inset to Fig. \ref{vida},
resulting in a novel MD regime at low-$T$ not observed in equilibrium. This
is confirmed by direct inspection of escape configurations, e.g. Fig. \ref{DSP0}.

One may get further physical insight by rewriting the rate (\ref{rate}) as 
$p+(1-p)\Psi (x_i)\equiv \Psi (x_i^{\text{ef}})$, with $x_i=\beta \Delta \mathcal{H}_{i}$ and 
$x_i^{\text{ef}}=\beta_{\text{ef}} \Delta \mathcal{H}_{i}$. The resulting effective temperature 
$T_{\text{ef}}(x_i,p)\equiv \beta _{\text{ef}}^{-1}$ then follows as 
$T_{\text{ef}}/T=x_i \, \left(x_i + \ln\left[ (1-p)/(1+p\text{e}^{x_i})\right]\right)^{-1}$. 
For any $p>0$, $T_{\text{ef}}$ changes from spin to spin and depends on
the \emph{local order} (e.g. number of broken bonds for a given spin): the smaller the
number of broken bonds, the higher the local order, and the larger the effective temperature.
Therefore, for $p>0$, the strength of fluctuations
affecting a spin increases with the local order parameter. This is the
fingerprint of \emph{multiplicative noise}, and it allows us to write a
Langevin equation which captures the essential physics. In its simplest,
0--dimensional form, this equation is $\partial _{t}\psi =\psi -\psi
^{3}+h+\sqrt{D+\mu \psi ^{2}} \, \xi (t)$, where $\xi (t)$ is a Gaussian white
noise with $\langle \xi (t)\rangle =0$ and $\langle \xi (t)\xi (t^{\prime
})\rangle =2\delta (t-t^{\prime })$, $D$ is the strength of the thermal noise, $h$
is a magnetic field, and $\mu $ is the renormalized version of the
nonequilibrium parameter $p$. This equation describes a \emph{Brownian
particle} in an asymmetric bimodal potential, $V(\psi )=-\frac{1}{2}\psi
^{2}+\frac{1}{4}\psi ^{4}-h\psi $, subject to fluctuations which increase
with $\psi ^{2}$ and whose amplitude remains non-zero as $D\rightarrow 0$ for
any $\mu >0$ \cite{large}. A full description of the problem, including the compact 
excitations observed in simulations, would of course involve the 
spatially-extended version of this equation. However, the above toy mean-field equation
already contains the essential competition between thermal ($D$) and nonequilibrium 
($\mu$) fluctuations in a metastable potential which characterizes our system.
The steady distribution of the stochastically--equivalent Fokker--Planck equation in 
Stratonovich sense is 
$P_{\text{st}}(\psi )=\Lambda \lbrack 2\sqrt{\mu (D+\mu \psi ^{2})}]^{-1}%
\text{exp}[-D^{-1}W(\psi )]$, where $W(\psi )=\frac{d}{2}\left[
\psi ^{2}-\left( 1+d\right) \ln \left( d+\psi ^{2}\right) \right] -hd^{1/2}\tan ^{-1}\left( \psi d^{-1/2}\right)$, 
with $d\equiv D/\mu $ and $\Lambda $ a normalization
constant. The extrema of the effective potential $W(\psi )$ are the same
than those of $V(\psi )$, namely $\psi _{k}=2\cos (\theta _{k})/\sqrt{3}$,
with $\theta _{k}=\frac{1}{3}[\cos ^{-1}(-\frac{1}{2}\sqrt{27}h)+2k\pi ]$,
and $k=0,1,2$. For $h<0$, $\psi _{0}$, $\psi _{1}$ and $\psi _{2}$
correspond to the metastable, stable and unstable extrema, respectively, and
the escape time from the metastable minimum is $\tau (D,\mu ,h)\approx 2\pi
\lbrack |V^{\prime \prime }(\psi _{0})V^{\prime \prime }(\psi _{2})|]^{-1/2}%
\text{exp}[\frac{1}{D}(W(\psi _{2})-W(\psi _{0}))]$, where $V^{\prime \prime
}=\partial _{x}^{2}V(x)$. Identifying $D$ with temperature, this approach
recovers the thermal NES phenomenon for $\tau $ observed in the microscopic
model (see bottom inset to Fig. \ref{vida}).

Summing up, we have shown that introducing a nonequilibrium condition has
important effects on metastable dynamics. The non--linear interplay between 
thermal and nonequilibrium fluctuations, as captured by $\sigma _{0}$, results in 
resonant low-$T$ phenomena, e.g., noise--enhanced stabilization of the metastable state 
and resonant domain wall propagation, that are not observed in equilibrium, but 
are likely to characterize a broad class of actual systems dominated by dynamic 
impurities. Surprisingly, these far-from-equilibrium phenomena can be understood
using nucleation theory, based on a nonequilibrium potential at a mesoscopic/cluster
level. This is possible because the relevant excitations driving the metastable-stable 
transition result from a competition between a surface and a bulk. This suggests that our 
approach may prove useful for studying metastability in many other nonequilibrium systems 
where a surface-bulk competition plays a significant role.

We acknowledge useful discussions with M.A Mu\~{n}oz, and financial support
from MEyC, FEDER, EU, and \textit{Junta de Andaluc\'{\i}a, }projects
FIS2005-00791, HPRN-CT-2002-00307 (DYGLAGEMEM) and FQM--165.


\begin{thebibliography}{99}
\bibitem{sigma} Low-- and high--temperature expansions of $\sigma _{0}(T,p)$
yield $\sigma _{0}\approx \lbrack T/T_{\text{ef}}^{\text{(I)}}]\sigma _{%
\text{e}}$, where $T_{\text{ef}}^{\text{(I)}}(T,p)$ is interpreted as an
effective interfacial temperature, and $\sigma _{\text{e}}(T)$ is the exact
equilibrium surface tension. At high$-T$, where thermal fluctuations
dominate over the nonequilibrium noise, $T_{\text{ef}}^{\text{(I)}}\propto T,
$ and $\sigma _{0}\propto \sigma _{\text{e}},$ which decreases with $T.$ On
the other hand, $T_{\text{ef}}^{\text{(I)}}$ saturates to a $T$-independent,
non-zero constant as $T\rightarrow 0$, due to the non-vanishing
nonequilibrium noise, so $\sigma _{0}$ increases linearly with $T$ at low--
$T$. This yields a non-monotonous $T$-dependence of $\sigma _{0}$ with a
maximum which agrees with the crossover in $T_{\text{ef}}^{\text{(I)%
}}$ \cite{Pablo2}.
\end{thebibliography}
\end{document}